\documentclass[aps,a4paper,prb,eqsecnum,showpacs,floatfix,twocolumn,merge,longbibliography,elide]{revtex4-1}

\usepackage{amsmath}
\usepackage{amsfonts}		
\usepackage[usenames,dvipsnames,svgnames]{pstricks}		
\usepackage{graphicx}		
\usepackage{tikz}			
\usepackage{tkz-graph}		

\usepackage{fouriernc}

\newcommand{\REM}[1]{}

\usepackage[%
unicode=true,%
colorlinks=true,%
linkcolor=blue,anchorcolor=blue,%
breaklinks=true,%
citecolor=blue,filecolor=cyan,%
menucolor=blue,%
breaklinks=true,%
urlcolor=blue]{hyperref}

\begin{document}
\title{Magnetic excitations in molecular magnets with complex bridges: \\
The tetrahedral molecule Ni$_4$Mo$_{12}$}
\author{M.~Georgiev}
\email{mgeorgiev@issp.bas.bg}
\affiliation{Institute of Solid State Physics, Bulgarian Academy of Sciences,
	Tsarigradsko Chauss\'ee 72, 1784 Sofia, Bulgaria}
\author{H.~Chamati}
\affiliation{Institute of Solid State Physics, Bulgarian Academy of Sciences,
	Tsarigradsko Chauss\'ee 72, 1784 Sofia, Bulgaria}

\date{\today}

\begin{abstract}
We investigate the spectroscopic magnetic
excitations in molecular magnets with complex intermediate structure
among the magnetic ions.
Our approach consists in introducing a modified spin Hamiltonian that allows
for discrete
coupling parameters accounting for all energetically favorable spatial distributions of the
valence electrons along the exchange bridges connecting 
the constituent magnetic ions. We discuss the physical relevance of
the constructed Hamiltonian and derive its eigenvalues.
The model is applied to explore the magnetic excitations of the
tetrameric molecular magnet Ni$_4$Mo$_{12}$. 
Our results are in a very good
agreement with the available experimental data.
We show that the experimental magnetic excitations in the named tetramer can be traced back to the
specific geometry and complex chemical structure of the exchange bridges
leading to the splitting and broadness of the peaks centered about 0.5
meV and 1.7 meV.
\end{abstract}

\pacs{75.10.-b, 75.10.Jm, 75.30.Et, 75.50.-y, 75.50.Xx}
\maketitle
\section{Introduction}\label{intro}
The physical properties, such as energy spectra, susceptibility,
etc., of magnetic clusters at the nanoscale depend on their size, 
shape and the presence of different bondings among the
constituent chemical elements and thus the distribution of ligands
between the magnetic ions (for more details see
\cite{roduner2006,chamati_theory_2013,sellmyer_novel_2015,sieklucka_molecular_2017}
and references therein).
Some prominent examples are Mn based magnetic compounds 
\cite{goodenough_1955,defotis_1990,law_2000,han_2004,perks_2012,gupta_2016,hanninen_2018}
and spin clusters with Ni magnetic ions 
\cite{loose_2008,panja_2017,das_2017,woods_2017}.

Magnetic molecules possess unique properties that can be
characterized with great accuracy both experimentally and
theoretically. Thus they are ideal tools
to gain useful insights into the underlying coupling mechanisms. 
The study of basic units like tetramers 
\cite{klemm_single-ion_2008,hubner_2017} prove the importance
of analytical methods in revealing the role of electrons correlations 
underpinning molecular magnetism.
On the experimental side, Inelastic Neutron Scattering (INS) 
\cite{lovesey_1986,malcolm_neutron_1989,furrer_neutron_2009,toperverg_neutron_2015} 
plays a central role in determining the relevant magnetic spectra. 
In complement to different magnetic measurement methods, INS technique appears
to be essential, and in the past decades it has been widely
applied to explore the properties of spin clusters. 
Experiments on the spin dimer
[Ni$_2$(ND$_2$C$_2$H$_4$ND$_2$)$_4$Br$_2$]Br$_2$
have demonstrated the important contribution of neutron spectroscopy \cite{stebler_intra_1982}. 
INS measurements were obtained for different magnetic clusters, such as: 
The trimer La$_4$Cu$_3$MoO$_{12}$, with strong intratrimer antiferromagnetic
interactions, where the copper ions form an isolated triangle
\cite{azuma_antiferromagnetic_2000}, the Fe based molecular wheel with eighteen 
spin-$\tfrac52$ ions \cite{ummethum_2012}, the dimer
SrCu$_2$(BO$_3$)$_2$ with observed multiplet excitations
\cite{kageyama_direct_2000,gaulin_high-resolution_2004}, the
polyoxomolybdate Mn$_{72}$Fe$_{32}$
\cite{garlea_probing_2006}, and the magnetic molecule Fe$_9$ in the presence
of an external magnetic field \cite{vaknin_magnetic_2014}.

In the present work we propose an approach based on the assumption that 
in a molecular magnet with multiple exchange bridges between any two magnetic 
centers (ions) the distribution of unpaired valence electrons is not unique.
Thus, the electron's density distribution might vary among the existing exchange bridges
affecting the transition energy, and consequently leading to either a broadened excitation 
width or splitting in the energy spectrum.
Accordingly the number of all energy levels form a set
that can utterly identify the most relevant spin bonds, despite being
indistinguishable.
To this end, we introduce a modified microscopic spin Hamiltonian with discrete couplings 
that incorporates distinct spin coupling mechanisms among equivalent spins 
allowing one to identify the different exchange paths. 
The proposed Hamiltonian leads to an unperturbed energy structure that distinguish the 
relevant magnetic features from those arising due the magnetic
anisotropy. It is worth mentioning that the present approach was successfully applied to the study
of the magnetic excitations in the trimers A$_3$Cu$_3$(PO$_4$)$_4$
with (A = Ca, Sr, Pb) \cite{georgiev_trimer_2019}.

\begin{figure}
		\includegraphics{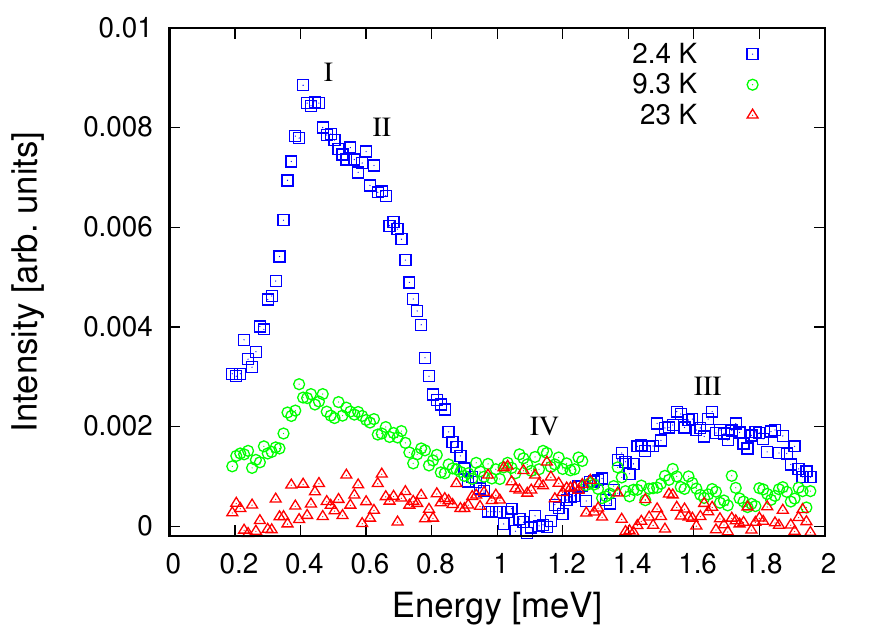}
	\caption{Background corrected data for the INS spectrum of the polycrystalline 
		Ni$_4$Mo$_{12}$ taken from Ref. \cite{nehrkorn_inelastic_2010}.
	}
	\label{fig:NiExp}       
\end{figure}

Here, the present method will be validated by reproducing
the experimentally
obtained INS spectrum of a molecular magnet that has generated a great
deal of interest among researchers both on the theoretical, as well as
the experimental sides. This is the magnetic molecule
[Mo$_{12}$O$_{30}$($\mu_2$-OH)$_{10}$H$_2$(Ni(H$_2$O)$_3$)$_4$], denoted by
Ni$_4$Mo$_{12}$, where four spin-1 Ni$^{2+}$ ions are
sitting on the vertices of a distorted tetrahedron \cite{muller_2000}.
We would like to point out that previous studies
\cite{nehrkorn_inelastic_2010,furrer_magnetic_PRB_2010} did not
succeed to obtain the main peaks and the broadening in the INS
spectrum, see FIG. \ref{fig:NiExp}, despite including
single-ion anisotropy or higher-order terms in the
Heisenberg model with nearest-neighbor interaction or even
using the Hubbard model.

\section{The model and the method}\label{sec:fundamprinc}
In order to identify the experimentally observed magnetic peaks, see FIG. \ref{fig:NiExp}, one has
to calculate the scattering intensities $I_{n'n}(\mathbf{q})$,
of the existing transitions and analyse their
dependence on the temperature and the magnitude of the neutron scattering vector.
For identical magnetic ions, we have
\cite{lovesey_1986,malcolm_neutron_1989,furrer_neutron_2009,toperverg_neutron_2015}
\begin{equation}\label{eq:ScatteringIntensities}
	I_{n'n}(\mathbf{q}) \propto F^2(\mathbf{q}) \sum_{\alpha, \beta} \varTheta^{\alpha \beta}
	S^{\alpha \beta} (\mathbf{q},\omega_{n'n}).
\end{equation}
Here $\omega_{n'n}$ is the transition frequency, 
$\mathbf{q}$ is the scattering vector, $F(\mathbf{q})$ is the
spin magnetic form factor \cite{jensen_rare_1991},
$\varTheta^{\alpha \beta}$ are the elements of polarization factor, 
$S^{\alpha \beta} (\mathbf{q},\omega_{n'n})$ are the magnetic
scattering functions and $\alpha,\beta,\gamma\in\{x,y,z\}$. 

To determine
the energy level structure and the transitions corresponding to
the experimentally observed magnetic spectra one needs a minimal
number of parameters to account for all couplings in the system. 
The principal assumption is that
the magnetic excitations of spin clusters obtained via INS are mainly governed by the
exchange of electrons between
effective spin magnetic centers. Then, the experimental data are interpreted in terms
of a well defined microscopic spin model.
In the absence of anisotropy, i.e. negligible spin-orbit coupling, 
the exchange interaction in molecular magnets can be described by 
the Heisenberg model
\begin{equation}\label{eq:HeisenbergHamiltonian}
	\hat{H} = \sum\limits_{i \ne j}^{} J_{ij}
	\hat{\mathbf{s}}_i \cdot \hat{\mathbf{s}}_j,
\end{equation}	
where $J_{ij} = J_{ji}$ is the exchange coupling that effectively accounts
for the exchange
interaction between the $i$th and $j$th ions.
However, to distinguish all magnetic excitations one has to use an appropriate
spin model leading to an energy sequence such that the 
scattering functions in \eqref{eq:ScatteringIntensities} 
identifies the relevant spin bonds.
Therefore, even with a selected \textit{a priori} spin coupling scheme
for a cluster with complex intermediate structure Hamiltonian 
\eqref{eq:HeisenbergHamiltonian} may not be fully adequate to 
obtain the correct energy structure.

\subsection{The effective spin Hamiltonian} \label{sec:spinmodel}
In order to characterize uniquely
each pair of magnetic centers in a magnetic cluster, 
bonded via more than one intermediate bridge,
we propose the following Hamiltonian
\begin{equation}\label{eq:AddHamiltonian}
	\hat{\mathcal{H}} = 
	\sum\limits_{i \ne j}^{} J_{ij} 
	\hat{\boldsymbol{\sigma}}_i \cdot \hat{\mathbf{s}}_j,
\end{equation}
where the couplings $J_{ij} = J_{ji}$
are effective exchange constants and the operator
$\hat{\boldsymbol{\sigma}}_i =
(\hat{\sigma}^x_i, \hat{\sigma}^y_i, \hat{\sigma}^z_i)$ 
accounts for the variation in valence electron's distribution
with respect to the $i$th magnetic center.

A detailed derivation of the
effective Hamiltonian \eqref{eq:AddHamiltonian} is very lengthy and falls 
beyond the scope the this study (the case of magnetic dimers is
published in Ref. \cite{georgiev_2018a}). In the following we will give
a brief account of the main steps leading to construct the ensuing
Hamiltonian \eqref{eq:AddHamiltonian}. Let us point out that our computations are based on the Molecular Orbital
Theory \cite{fleming_molecular_2009} in terms of the 
complete active space self consistent field method
\cite{roos_complete_2007,szalay_multicon_2012}.
We assume that each exchange bridge connecting two magnetic centers possess 
a number of paired valence electrons, nuclei and thus favor a particular
spatial distributions of electrons.
The canonical Hamiltonian leading to \eqref{eq:AddHamiltonian} accounts for the 
kinetic energy, electron-electron and electron-nuclei interactions of all 
valence electrons within the adiabatic approximation.
The electrons are considered as delocalized, occupying molecular orbitals 
$\phi_{k,m_i}(\mathbf{r}_i)$, $k\in\mathbb{N}$,
given by linear combinations of atomic orbitals 
$\psi^{\eta}_{\mu_{\eta i},m_i}(\mathbf{r}_i)$, where $\mathbf{r}_i$ are the coordinates of the
$i$th electron, $\mu_{\eta i}$ label the electronic shell and subshell with respect to the 
$i$th electron and $\eta$ nucleus, $m_i$ is the spin magnetic quantum number of the $i$th electron.
The state function of each electron configuration is given by a linear combination of
Slater determinants (of the orbitals
$\phi_{k,m_i}(\mathbf{r}_i)${}), where the 
symmetrization is performed with respect to the spin quantum numbers $s_{ij}$ 
of all electron pairs. The corresponding basis set gives the number of all 
probable electrons distributions along all exchange bridges.
In the simplest case of $N$ valence electrons with $N-2$ pairs, one of the 
basis states is written as
\begin{align}\label{eq:TwoOrbital}
	\Psi^{k,k'}_{S,M}(\mathbf{r}_1,\ldots,\mathbf{r}_{N})\equiv
	& 
	\!\!\!
	\sum_{P_{\mathbf{r}_1\ldots\mathbf{r}_{N}}}
	\!\!\!
	c_{\mathbf{r}_{N-1},\mathbf{r}_{N}}
	\!\!\!
	\prod_{i}^{\frac{N}{2}-1}
	\!\!
	\frac{
		\Phi^i_{s_{2i-1,2i}}(\mathbf{r}_{2i-1},\mathbf{r}_{2i}) 
	}{\sqrt{2^{\frac{N}{2}}N!}}
	\nonumber \\ 
	& 	\times
	\Psi^{k,k'}_{s_{N-1,N}}(\mathbf{r}_{N-1},\mathbf{r}_{N})
	\lvert S,M\rangle,
\end{align}
where the sum runs over all permutations on the set of
coordinates $\mathbf{r}_1\ldots\mathbf{r}_{N}$. 
Both unpaired electrons occupy $k$th and $k'$th molecular orbitals with
\begin{align}
	\Psi^{k,k'}_{s_{N-1,N}}(\mathbf{r}_{N-1},\mathbf{r}_{N})=
	&
	\tfrac{1}{\sqrt{2}}
	\big[ 
	\phi_{k}(\mathbf{r}_{N-1})\phi_{k'}(\mathbf{r}_{N})+
	\nonumber \\ \nonumber
	& (-1)^{s_{N-1,N}}
	\phi_{k'}(\mathbf{r}_{N-1})\phi_{k}(\mathbf{r}_{N})	
	\big].
\end{align}
The remaining, paired electrons, are described by the states
\begin{align}
	\Phi^i_{s_{2i-1,2i}}(\mathbf{r}_{2i-1},\mathbf{r}_{2i})=
	&
	\tfrac12 
	\big[ 
	\phi_i(\mathbf{r}_{2i-1})\phi_i(\mathbf{r}_{2i})+
	\nonumber \\ \nonumber
	&	(-1)^{s_{2i-1,2i}}
	\phi_i(\mathbf{r}_{2i-1})\phi_i(\mathbf{r}_{2i})
	\big].
\end{align}
Moreover, for $i,j = 1,\ldots, N$ the permutation coefficients 
$$
c_{\mathbf{r}_{i},\mathbf{r}_{j}}= (-1)^{i+j+1},
\quad
c_{\mathbf{r}_{j},\mathbf{r}_{i}}=(-1)^{s_{N-1,N}}(-1)^{i+j+1},
$$
account for the antisymmetry of \eqref{eq:TwoOrbital} and
the spin part is given by
$$
\lvert S,M\rangle=\otimes^{\frac{N}{2}}_{i=1}
\lvert s_{2i-1,2i},m_{2i-1,2i}\rangle.
$$
The effective Hamiltonian accounting for all possible configurations
of interactions
is built according to the expectation values of the initial Hamiltonian representing
all electron-electron and electron-nuclei interactions. 
The effective state associated with \eqref{eq:TwoOrbital} is represented by 
$\big\lvert\Psi^{\tau}_{S,M}\big\rangle$, where $\tau=(k,k')$.
The generic wave function describing the multiple bridged structure includes all
probabilities related with the spatial distribution of unpaired electrons. 
Thus, we may compute
the eigenstates of the ensuing spin Hamiltonian in terms of an appropriate linear
combination of the effective states.
In the case of \eqref{eq:TwoOrbital}, this is given by the superposition
\begin{equation}\label{eq:AllPathsStatesTriplet}
	\lvert s,m\rangle
	=\sum_{\tau}c^\tau_{S} 
	\big\lvert\Psi^{\tau}_{S,M}\big\rangle,
\end{equation}
where 
$\tau$ runs over the number of all existing electrons' configurations
that depend on all intermediate bridges and
$s=0,1$, $m=0,\pm1$ are the corresponding effective spin 
and magnetic quantum numbers, that obey the conservation of angular momentum 
$S\equiv s$, $M\equiv m$. Further, $c^\tau_{S}$ is the associated probability
coefficient that depends on $S$.
The expectation values of the canonical Hamiltonian are a part of the eigenstates
relevant to \eqref{eq:AllPathsStatesTriplet}.
Within the framework of the given example they are represented by the sum
\begin{equation}\label{eq:EffEnergies}
E_{s,m}=\sum_{\tau} \big|c^\tau_{S}\big|^2E^{\tau}_{S,M}.
\end{equation}
The elements in the last sum are functions of the Coulomb 
$U_\tau, V_\tau$, hopping 
$t_\tau$ and direct exchange integrals $D_\tau$ relevant to each intermediate bridge. 
For example, 
$$
E^\tau_{1,M}=V_\tau-D_\tau,
\qquad
M=0,\pm1,
$$
$$
E^\tau_{0,0}= D_\tau+\frac{U_\tau+V_\tau}{2}-
\sqrt{4t^2_\tau+\frac{\left(U_\tau-V_\tau\right)^2}{4}}.
$$
Within the effective spin space, determined by $s$ and $m$, one can account for only one
transition with energy $|\Delta E|=|E_{1,m}-E_{0,0}|$ due to the 
$3$-fold degeneracy of the triplet level. However, for different
spatial distributions of the considered electrons the values of \eqref{eq:EffEnergies}
alter and accordingly the energy of the transition changes.
Such effect is not related neither to the magnetic anisotropy nor higher order multiple interactions.
Therefore, conventional bilinear spin Hamiltonians with only one exchange coupling
and additional interacting terms is not able to account for the variation in $\Delta E$.

In order to address these features we introduce Hamiltonian 
\eqref{eq:AddHamiltonian} that depends upon the parameters described hereafter.

\subsection{Properties of the $\hat{\boldsymbol{\sigma}}$--operators}
For a single spin the square and the $z$ component of each operator
$\hat{\boldsymbol\sigma}$ are completely determined in the
basis of the total spin component $\hat{s}^z$, such that for all $i$ and 
$\alpha \in \{x,y,z\}$
\begin{equation}\label{eq:Sigma_i}
	\hat{\sigma}^\alpha_i \lvert \ldots ,s_i,m_i, \ldots \rangle
	=
	a^{s_i,m_i}_i \hat{s}^{\alpha_{\phantom{i}}}_i 
	\lvert \ldots ,s_i,m_i, \ldots \rangle,
\end{equation}
where $a^{s_i,m_i}_i \in \mathbb{R}$. 
Furthermore, the $\sigma$ rising and lowering operators
obey the equations
\begin{equation}\label{eq:Sigma_LR_i}
	\hat{\sigma}^{\pm}_i \lvert \ldots ,s_i,m_i, \ldots \rangle
	=
	a^{s_i,m_i}_i \hat{s}^{{\pm}_{\phantom{i}}}_i 
	\lvert \ldots ,s_i,m_i, \ldots \rangle.
\end{equation}
For all $i$, the square of $\sigma_i$ commutes
only with its $z$ component. 
Its eigenvalues depend on $m_i$ and according to
\eqref{eq:Sigma_i} and \eqref{eq:Sigma_LR_i} 
one can distinguish two cases:
(1) $m_i=\pm s_i$; (2) $-s_i<m_i<s_i$, 
where $s_i\ne0$, with the respective eigenvalues
\begin{subequations}\label{eq:SigmaSquareEigenvalue_i}
	\begin{equation}\label{eq:SigmaSquareEigenvalue_mi=si}
		\left( a^{s_i,\pm s_i}_i\right)^2 s^2_i+a^{s_i,\pm s_i}_i 
		a^{s_i,\pm(s_i-1)}_i s_i,
	\end{equation}
	\begin{align}\label{eq:SigmaSquareEigenvalue_mi}
		& \tfrac12 a^{s_i,m_i}_i \left[a^{s_i,m_i+1}_i + 
		a^{s_i,m_i-1}_i\right]s_i(s_i+1)
		+\left( a^{s_i,m_i}_i\right)^2 m^2_i
		\nonumber \\
		& - \tfrac12 a^{s_i,m_i}_i m_i \left[a^{s_i,m_i+1}_i(m_i+1) 
		+ a^{s_i,m_i-1}_i (m_i-1)\right].
	\end{align}
\end{subequations}

On the other hand when the spins of $i$th and $j$th magnetic ions are
coupled, with total spin operator
$\hat{\mathbf{s}}_{ij}=\hat{\mathbf{s}}_{i}+\hat{\mathbf{s}}_{j}$,
the relation \eqref{eq:Sigma_i} enters a more general and complex expression.
To explore the properties of the coupled spins one has to
work with the total $\sigma$-operator $\hat{\boldsymbol{\sigma}}_{ij}$.
Its $z$ component and square are completely determined in the basis of 
the spin operator $\hat{\mathbf{s}}^2_{ij}$.
Similar to eq. \eqref{eq:Sigma_i} for all $i\ne j$ and 
$\alpha \in \{x,y,z\}$, we have
\begin{equation}\label{eq:Sigma_ij}
	\hat{\sigma}^\alpha_{ij} \lvert \ldots ,s_{ij},m_{ij}, \ldots \rangle
	=
	a^{s_{ij},m_{ij}}_{ij} \hat{s}^{\alpha_{\phantom{j}}}_{ij} 
	\lvert \ldots ,s_{ij},m_{ij}, \ldots \rangle,
\end{equation}
where $a^{s_{ij},m_{ij}}_{ij} \in \mathbb{R}$. 
The corresponding rising and lowering operators obey
\begin{equation}\label{eq:Sigma_LR_ij}
	\hat{\sigma}^{\pm}_{ij} \lvert \ldots ,s_{ij},m_{ij}, \ldots \rangle
	=
	a^{s_{ij},m_{ij}}_{ij} \hat{s}^{{\pm}_{\phantom{i}}}_{ij} 
	\lvert \ldots ,s_{ij},m_{ij}, \ldots \rangle.
\end{equation}
The eigenvalues of $\hat{\boldsymbol{\sigma}}^2_{ij}$ depend on $m_{ij}$.
Therefore having in mind the following two cases $m_{ij}=\pm s_{ij}$,
and $-s_{ij}<m_{ij}<s_{ij}$, where $s_{ij}\ne0$, the eigenvalues read
\begin{subequations}\label{eq:SigmaSquareEigenvalue_ij}
	\begin{equation}\label{eq:SigmaSquareEigenvalue_mij=sij}
		\left( a^{s_{ij},\pm s_{ij}}_{ij}\right)^2 s^2_{ij}+a^{s_{ij},
			\pm s_{ij}}_{ij} a^{s_{ij},\pm(s_{ij}-1)}_{ij} s_{ij},
	\end{equation}
	\begin{align}\label{eq:SigmaSquareEigenvalue_mij}
		& \tfrac12 a^{s_{ij},m_{ij}}_{ij} \left[a^{s_{ij},m_{ij}+1}_{ij} + 
		a^{s_{ij},m_{ij}-1}_{ij}\right]s_{ij}(s_{ij}+1)
		\nonumber \\
		& +\left( a^{s_{ij},m_{ij}}_{ij}\right)^2 m^2_{ij}
		- \tfrac12 a^{s_{ij},m_{ij}}_{ij} m_{ij} 
		\nonumber \\
		& \times\left[a^{s_{ij},m_{ij}+1}_{ij}(m_{ij}+1) 
		+ a^{s_{ij},m_{ij}-1}_{ij} (m_{ij}-1)\right].
	\end{align}
\end{subequations}
The corresponding $\sigma$-operators share a single
coefficient and for $i \ne j$ and $\alpha \in \{x,y,z\}$, 
we have
\begin{equation}\label{eq:Sigma_k}
	\hat{\sigma}^{\alpha}_i \lvert \ldots ,s_{ij},m_{ij}, \ldots \rangle
	=
	a^{s_{ij},m_{ij}}_{ij} \hat{s}^{\alpha_{\phantom{j}}}_i 
	\lvert \ldots ,s_{ij},m_{ij}, \ldots \rangle.
\end{equation}

We further assume that
the $\sigma$-operators preserve the corresponding
spin magnetic moment and for a non coupled spin
obey the following constraints
\begin{subequations}\label{eq:Constraint_i}
	\begin{equation}\label{eq:SigmaZ_i}
		\hat{\sigma}^z_i \lvert \ldots ,s_i,m_i, \ldots \rangle
		=
		m_i \lvert \ldots ,s_i,m_i, \ldots \rangle,
	\end{equation}
	\begin{equation}\label{eq:SigmaSquare_i}
		\hat{\boldsymbol{\sigma}}^2_i \lvert \ldots ,s_i,m_i, \ldots \rangle
		=
		s_i(s_i+1) \lvert \ldots ,s_i,m_i, \ldots \rangle.
	\end{equation}
\end{subequations}
Similarly, when the $i$th and $j$th spins are 
coupled, for all $i \ne j$ we have
\begin{subequations}\label{eq:Constraint_ij}
	\begin{equation}\label{eq:SigmaZ_ij}
		\hat{\sigma}^z_{ij}
		\lvert\ldots,s_{ij},m_{ij},\ldots \rangle =
		m_{ij} \lvert\ldots,s_{ij},m_{ij},\ldots \rangle,
	\end{equation}
	\begin{equation}\label{eq:SigmaSquare_ij}
		\hat{\boldsymbol{\sigma}}^2_{ij} 
		\lvert\ldots,s_{ij},m_{ij},\ldots \rangle =
		s_{ij}(s_{ij}+1) 
		\lvert\ldots,s_{ij},m_{ij},\ldots \rangle.
	\end{equation}
\end{subequations}

Taking into account \eqref{eq:Constraint_i} together with
expressions \eqref{eq:SigmaSquareEigenvalue_mi} for all $m_i$ we have
$a^{s_i,m_i}_i=1$. 
Further, according to constraints \eqref{eq:Constraint_ij} and 
eqs. \eqref{eq:SigmaSquareEigenvalue_ij} we distinguish three cases:

\textbf{(1)} $s_{ij} \ne 0$, $m_{ij} \ne 0$:
Then, for all $m_{ij}$, $a^{s_{ij},m_{ij}}_{ij}=1.$
As a result the transformations of eigenvectors
via the $\sigma$-operator coincide with those defined by 
its corresponding spin operator.
Therefore, all couplings will be constants and the Hamiltonian
\eqref{eq:AddHamiltonian} will capture the same features as its
Heisenberg parent.

\textbf{(2)} $s_{ij} \ne 0$ and $m_{ij}=0$:
The corresponding coefficient cannot be determined from Eq.
\eqref{eq:SigmaZ_ij} and from eqs. \eqref{eq:SigmaSquareEigenvalue_mij} and
\eqref{eq:SigmaSquare_ij} one obtains
\begin{equation}\label{eq:as0_ij}
	a^{s_{ij},\pm1}_{ij}=a^{s_{ij},0}_{ij}=\pm1.
\end{equation}
We would like to point out that the ``minus''
sign is irrelevant to the case of two unpaired electrons.

\textbf{(3)} $s_{ij} = 0$: The associated parameter
remains unconstrained and there exist a
set of coefficients $c^{n}_{ij}\in\mathbb{R}$ $\forall n\in\mathbb{N}$, 
such that
\begin{equation}\label{eq:a00_ij}
	a^{0,0}_{ij} \in \{c^{n}_{ij}\}_{n\in\mathbb{N}}.
\end{equation} 

The values of $c^{n}_{ij}$
depend on the number of unpaired valence electrons and 
intermediate nonmagnetic ions of the respective exchange bridges.
Depending on the type of exchange these effective 
coefficients are functions of the Coulomb, hopping and 
exchange integrals.
Thereby, for a linear cluster with only one bonding anion between
magnetic cations and a unique electron's spatial distribution, one would obtain the limit
$|c^{n}_{ij}-c^{k}_{ij}| \to 0, \ \forall \ n \ne k$, 
where $c^{n}_{ij}\to1$. Accordingly,
the changes in electron's distribution could be considered as negligible pointing 
to sharpened peaks in the magnetic 	
spectrum.
On the other hand, the inequality $|c^{n}_{ij}-c^{k}_{ij}|>0$ for all 
$n \ne k$, would have to be considered as a sign 
for the presence of exchange paths of different energy,
\textit{i.e.} more than one favorable spatial distribution of unpaired electrons,
and therefore of increased
excitation width in energy.
As an example, if $i$th and $j$th magnetic centers are linked via more than one 
exchange bridge of complex chemical structure, 
then one may expect
that the exchange path is not unique. 
In such case according to \eqref{eq:AllPathsStatesTriplet} and 
\eqref{eq:EffEnergies} the transition energy 
reads
$$
|\Delta E_n|=\sum_{\tau} 
\left| \big|c^\tau_{n,1}\big|^2 E^{\tau}_{1,M}
-\big|c^\tau_{n,0}\big|^2 E^{\tau}_{0,0}\right|,
$$ 
where $n$ assigns a unique transition energy to a certain 
number of favorable spatial distributions. 
Hence, with Hamiltonian \eqref{eq:AddHamiltonian} and taking into account 
\eqref{eq:a00_ij} we can express all existing transitions by
\begin{equation}\label{eq:TranEnergy_a1}
	\Delta E_n=\tfrac12 J_{ij} (3c^{n}_{ij}\pm1),
	\qquad
	a^{1,0}_{ij}=\pm1.
\end{equation}
The set of values $\Delta E_n$ will correspond to a broadened
peak in the magnetic spectrum. Applying the relation $\Delta E_n=2J_{c^{n}_{ij}}$,
where $J_{c^{n}_{ij}}$ is the $n$th value of the Heisenberg type exchange coupling
from \eqref{eq:TranEnergy_a1}
we thus obtain
\begin{equation}\label{eq:cij_a1}
	c^{n}_{ij}= \frac{4}{3}\frac{J_{c^{n}_{ij}}}{J_{ij}}\mp\frac13,  
	\qquad
	a^{1,0}_{ij}=\pm 1.
\end{equation}
As we will see later this approach 
allows one to explain in details the experimentally observed splitting and 
broadening of magnetic spectra in the molecular magnet
Ni$_4$Mo$_{12}$, see e.g. FIG. \ref{fig:NiExp}.

\section{The tetramer N\lowercase{i}$_4$M\lowercase{o}$_{12}$}
\begin{figure}[!ht]
	\centering
\includegraphics[scale=0.8]{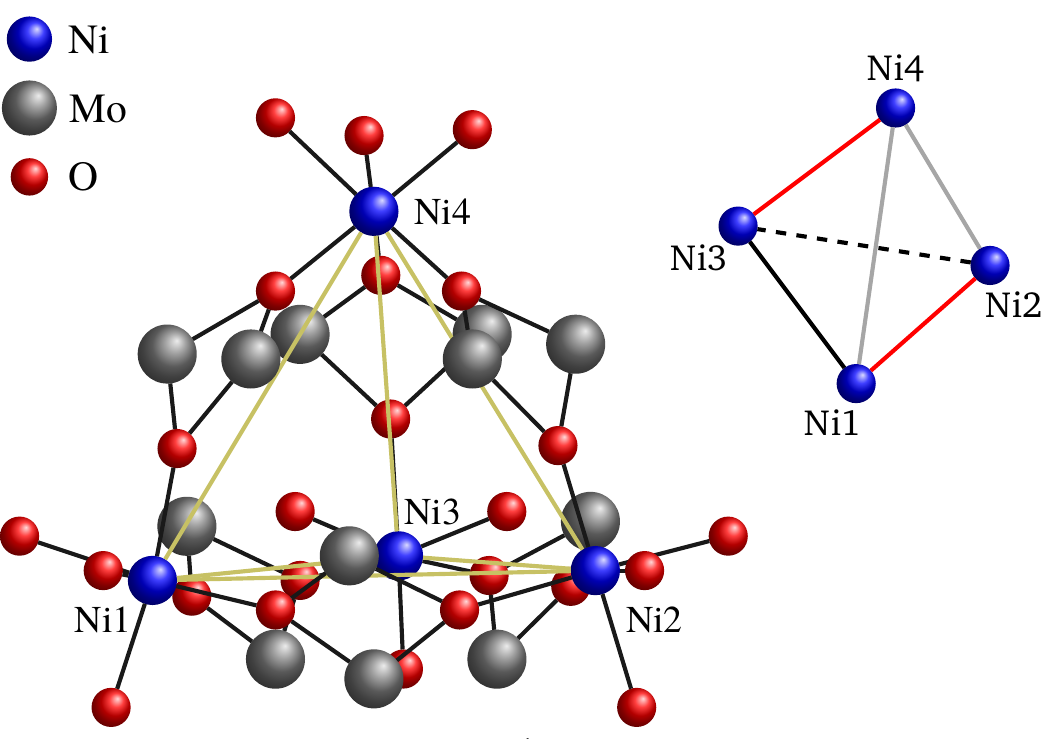}
	\caption{
		Sketch of the structure of the molecular
		nanomagnet Ni$_4$Mo$_{12}$. The inset represents a schematic view of 
		the arrangement of Ni ions (blue balls). The gray
		lines represent the two shorter distances, while the red lines show the
		effective spin-1 dimers. \label{fig:Nickel}}
\end{figure}

The indistinguishable spin-one $\mathrm{Ni^{2+}}$ ions of the spin
cluster compound $\mathrm{Ni_4Mo_{12}}$, are arranged
on the vertices of a distorted tetrahedron, see
FIG. \ref{fig:Nickel}. For this molecule 
the bonds Ni1-Ni2 and Ni3-Ni4
are slightly shorter about 0.03 \AA\ than the
other four  \cite{furrer_magnetic_PRB_2010}. Notice that the
intermediate bridges contain ions of molybdenum, oxygen and hydrogen.

To perform a thorough analysis of the magnetic excitations of the
compound $\mathrm{Ni_4Mo_{12}}$ obtained by INS experiments (see e.g.
FIG. \ref{fig:NiExp}) reported
in Ref. \cite{nehrkorn_inelastic_2010,furrer_magnetic_PRB_2010} we
employ Hamiltonian \eqref{eq:AddHamiltonian}.
Notice that the symmetry of the magnetic cluster imply
$J_{ij}=J$ and further we assume that
the magnetic centers Ni1-Ni2 and Ni3-Ni4 are coupled,
which defines these bonds as the intersection of two different planes.
Therefore, we have the total spin eigenstates
$\lvert s_{12},s_{34},s,m \rangle$,
four $\hat{\boldsymbol{\sigma}}$ operators for each constituent magnetic ion and two 
bond operators corresponding to both Ni1-Ni2 and Ni3-Ni4 spin pairs.

We point out that
$\hat{\boldsymbol{\sigma}}_1$ and $\hat{\boldsymbol{\sigma}}_2$ 
are related with the 
Ni1-Ni2 pair that share the coefficients 
$a^{s_{12},m_{12}}_{12}$ of the total bond
$\hat{\boldsymbol{\sigma}}_{12}$. The operators $\hat{\boldsymbol{\sigma}}_3$ 
and $\hat{\boldsymbol{\sigma}}_4$ are associated with the
coefficient $a^{s_{34},m_{34}}_{34}$ of
$\hat{\boldsymbol{\sigma}}_{34}$.
Consequently from \eqref{eq:AddHamiltonian} we obtain the Hamiltonian
\begin{align}\label{eq:NickelHamilton}
\hat{\mathcal{H}} =  & \ J 
\left( 
\hat{\boldsymbol{\sigma}}_1\cdot\hat{\mathbf{s}}_2 + 
\hat{\boldsymbol{\sigma}}_2\cdot\hat{\mathbf{s}}_1 +
\hat{\boldsymbol{\sigma}}_3\cdot\hat{\mathbf{s}}_4 +
\hat{\boldsymbol{\sigma}}_4\cdot\hat{\mathbf{s}}_3
\right) 
\nonumber \\ 
& + J 
\left( 
\hat{\boldsymbol{\sigma}}_{12}\cdot\hat{\mathbf{s}}_{34} +
\hat{\boldsymbol{\sigma}}_{34}\cdot\hat{\mathbf{s}}_{12}
\right).
\end{align}
With the applied effective spin-one spins the tetramer
exhibits in total eighty one eigenstates without taking into account the
quadrupolar, octupolar and other eigenfunctions related with higher
symmetries.
The ground state of this nanomagnet is a singlet with possible eigenstates
$\{\lvert 0,0,0,0 \rangle,\lvert 1,1,0,0 \rangle, \lvert 2,2,0,0 \rangle\}$.
On the other hand, the selection rules imply that the ground state
excitations must be related with singlet-triplet transitions and
since the quantum numbers $s_{14}$ and $s_{23}$ cannot be
simultaneously varied, we
deduce that the ground state is,
related to the formation of two local triplets, i.e. $s_{14}=1$ and $s_{23}=1$.
The triplet
eigenstates are eighteen. Those, three in total,
characterized by the local quintets $s_{14}=2$ and $s_{23}=2$
are not adequate to the
established selection rules and nine are directly connected
to experimental spectra.

\begin{figure*}[ht!]
\centering
\includegraphics[scale=1]{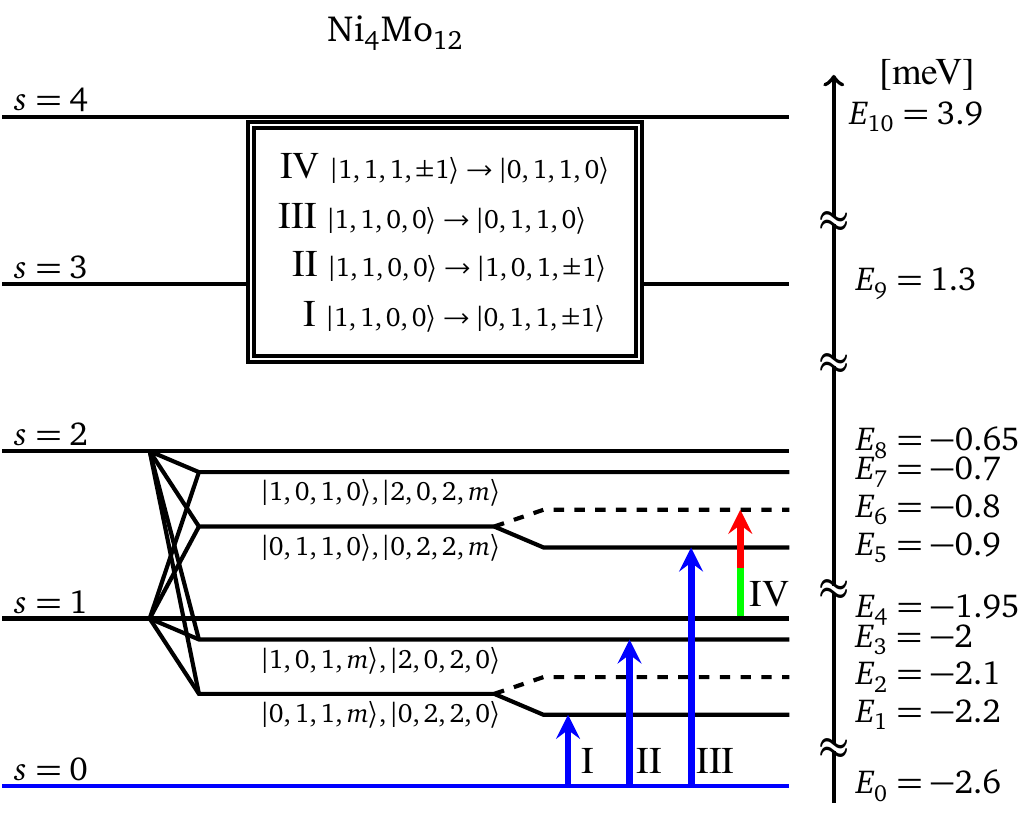}
	\caption{
		Energy level structure and the corresponding transitions of
		Ni$_4$Mo$_{12}$. The blue line and
		arrows stands for the ground state energy and the ground state excitations, respectively.
		The red arrow marks the excited transition and the corresponding initial
		level is shown in red. The dashed lines represent the centers of the two bands.
		All transitions are denoted with respect to the experimental data shown on 
		FIG. \ref{fig:NiExp} provided from Ref. \cite{nehrkorn_inelastic_2010}. 
	}
	\label{fig:NickelSpectra}
\end{figure*}

\subsection{Energy levels}
According to the selected coupling scheme we denote the
eigenvalues of \eqref{eq:NickelHamilton} by 
$E^{m}_{s_{12},s_{34},s}$.
The ground state is the singlet $\lvert 1,1,0,0 \rangle$. Therefore, using 
\eqref{eq:Constraint_ij} we get $a^{1,m_{12}}_{12}=a^{1,m_{34}}_{34}=1$
and taking into account \eqref{eq:NickelHamilton} we obtain
$$
E^{0}_{1,1,0} = -8J.
$$
With the triplet eigenstates $\lvert 0,1,1,m\rangle$, $m_{34}\equiv m=0,\pm1$, 
the spins of Ni1 and Ni2 ions are coupled in singlet, 
the parameter $a^{0,0}_{12}$ remains unconstrained 
and can be determined using INS experimental data.  
For the corresponding energies we get
\begin{equation*}
E^{m}_{0,1,1}= -2Ja^{1,m}_{34}-4Ja^{0,0}_{12}, \quad m=0, \pm 1,
\end{equation*}
where according to \eqref{eq:as0_ij} we have $a^{1,0}_{34}=\pm1$. 
Analyzing the Ni$_4$Mo$_{12}$ spectrum we further obtain
$a^{0,0}_{12}\in\{c^1_{12},c^2_{12}\}$.

When the singlet bond is Ni3-Ni4 the eigenstates are 
$\lvert 1,0,1,m\rangle$, $m_{12}\equiv m=0,\pm1$, 
the value of $a^{0,0}_{34}$ remains unconstrained 
and according to \eqref{eq:NickelHamilton} we have
\begin{equation*}
E^{m}_{1,0,1}= -2Ja^{1,m}_{12}-4Ja^{0,0}_{34}, \quad m=0, \pm 1,
\end{equation*}
where $a^{1,0}_{12}=\pm1$. Without loss of generality we set $a^{0,0}_{34}=c_{34}$.

For all remaining triplets, i.e. 
$\lvert 1,1,1,m\rangle$, $\lvert2,2,1,m\rangle$, $\lvert 2,1,1,m\rangle$ 
and $\lvert 1,2,1,m\rangle$, where $m=0,\pm 1$, the corresponding 
coefficient are constrained, $a^{s_{12},m_{12}}_{12}=1$ and 
$a^{s_{34},m_{23}}_{34}=1$. Thus, we obtain
$$
E^{m}_{1,1,1} = E^{m}_{2,2,1} =
E^{m}_{2,1,1} = E^{m}_{1,2,1} = -6J.
$$
The tetramer Ni$_4$Mo$_{12}$ exhibits also a singlet bond at the quintet level.
The energies associated with the Ni1-Ni2 singlet bond and eigenstates 
$\lvert 0,2,2,m \rangle$, where $m_{34}\equiv m$ are
\begin{equation*}
E^{m}_{0,2,2} = 2Ja^{2,m}_{34}-4Ja^{0,0}_{12}, \quad m=0, \pm 1,\pm 2,
\end{equation*}
where, the coefficients are determined by $a^{2,0}_{34}=\pm1$ and 
$a^{0,0}_{12}\in\{c^1_{12},c^2_{12}\}$. 

Similarly, if the spins of third and fourth ions are in a singlet state, 
where the corresponding eigenstates are 
$\lvert 2,0,2,m \rangle$, then the Hamiltonian in \eqref{eq:NickelHamilton}
yield the following energy values
\begin{equation*}
E^{m}_{2,0,2} = 2Ja^{2,m}_{12}-4Ja^{0,0}_{34}, \quad m=0, \pm 1,\pm 2,
\end{equation*}
where $a^{2,0}_{12}=\pm1$.

With respect to the other twelve quintet eigenstates the coefficients
$a^{s_{12},m_{12}}_{12}=a^{s_{34},m_{34}}_{34}=1$ and 
therefore,
$$
E^{m}_{2,2,2} = E^{m}_{1,1,2} =
E^{m}_{2,1,2} = E^{m}_{1,2,2} = -2J.
$$
For the remaining two levels we obtain 
$a^{s_{12},m_{12}}_{12}=1$ and $a^{s_{34},m_{34}}_{34}=1$.
The energy sequence follows the Land\'e interval rule $E_{s+1} -
E_{s} = 2Js$, see e.g. FIG. \ref{fig:NickelSpectra}. 
The septet level is twenty one fold degenerate and it is defined by the
eigenstates $\lvert 2,1,3,m \rangle, \lvert 1,2,3,m \rangle, \lvert
2,2,3,m \rangle$ with $m=0,\pm1,\pm2,\pm3$. All corresponding
energies have equal value
$$
E^{m}_{2,1,3} = E^{m}_{1,2,3} =
E^{m}_{2,2,3} = 4J.
$$
Applying the nonet state $\lvert 2,2,4,m \rangle$,
where $m=0,\ldots,\pm4$ we end up with
$E^{m}_{2,2,4} = 12J$.

The described energy level structure is illustrated on 
FIG. \ref{fig:NickelSpectra}.	
In what follows we find the following notations more convenient
\begin{equation*}
\begin{array}{lll}
E_0 = -8J, & E_1 = -2J-4Jc^1_{12}, \\
E_2 = -2J-4Jc^2_{12}, & E_3 = -2J-4Jc_{34}, \\ 
E_4=-6J,  & E_5 = 2J-4Jc^1_{12}, \\
E_6 = 2J-4Jc^2_{12}, & E_7 = 2J-4Jc_{34}, \\ 
E_8=-2J, & E_9=4J, \\ 
E_{10}=12J. & {}
\end{array}
\end{equation*}

\subsection{Scattering Intensities}

\begin{figure}[th!]
	\includegraphics[]{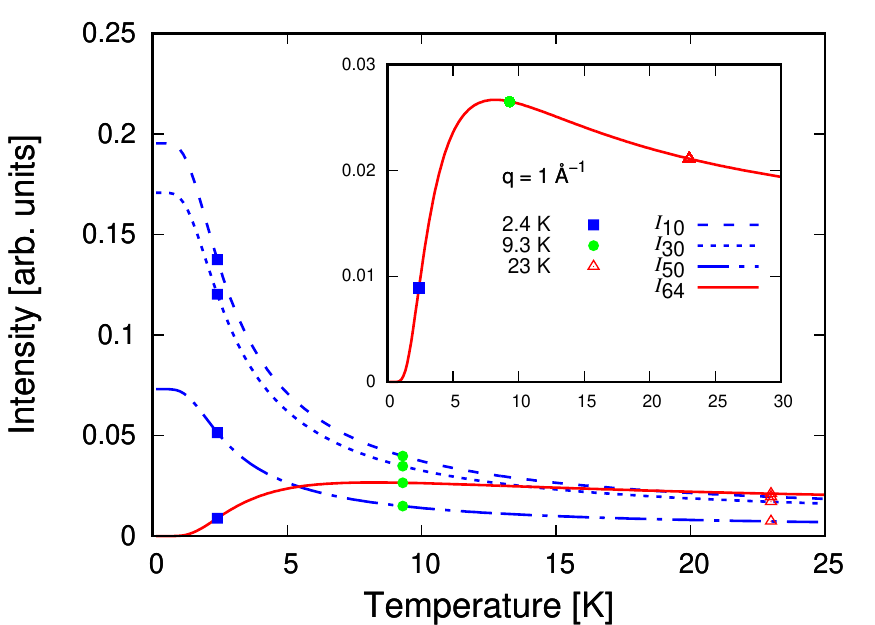}
	\caption{
		Intensities as a function of the absolute temperature. $I_{10}$, $I_{30}$
		and $I_{50}$ correspond to the ground state transitions with
		energies 0.4 meV, 0.6 meV and 1.7 meV, respectively. The
		intensity $I_{64}$ in the inset stands for the excited transition with
		energy 1.15 meV.
		The blue squares, the green circles and red triangles point to the
		values of intensities in table~\ref{tab:NickelIntensThreeTemp}.
	}
	\label{fig:NiIntensity}
\end{figure}

\begin{figure}[th!]
	\includegraphics[]{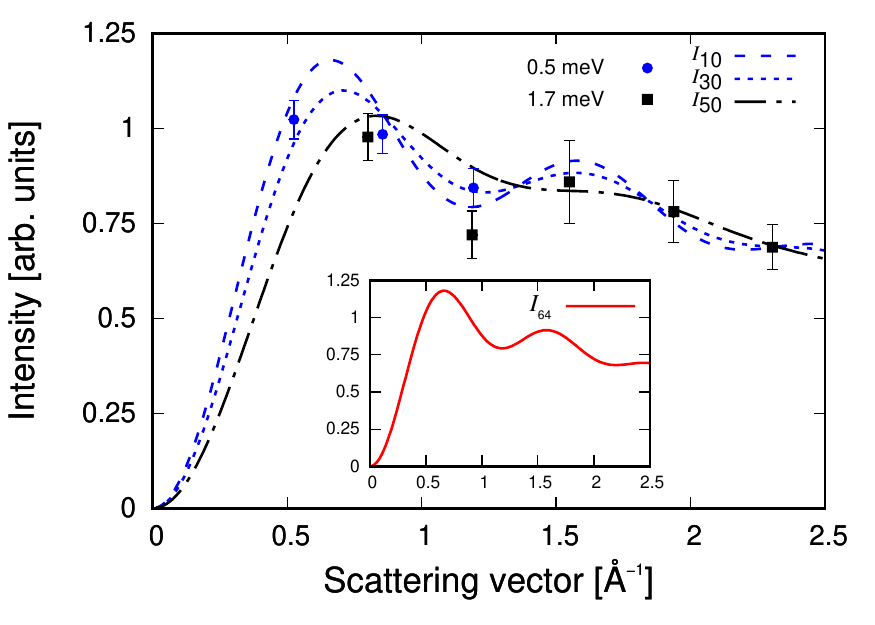}
	\caption{
		Normalized by $\gamma_{n'n}$ intensities as a function of the
		scattering vector along with the experimental data of 
		Ref. \cite{furrer_magnetic_PRB_2010}. $I_{10}$, $I_{30}$ and $I_{50}$ correspond
		to ground state transitions with energies 0.4 meV, 0.6 meV and
		1.7 meV, respectively. The intensity $I_{64}$ stands for the
		excited transition with energy 1.15 meV. The inset
		shows the intensity $I_{64}$ that coincides
		with the function $I_{10}$.
	}
	\label{fig:NiIntFFactor}
\end{figure}

The INS selection rules are $\Delta s = 0, \pm 1$, $\Delta
m = 0, \pm 1$ and $\Delta s_{12} = 0, \pm 1$, $\Delta s_{34} = 0, \pm
1$. Here the transitions $\Delta s_{12} \ne 0$
and $\Delta s_{34} \ne 0$ are not allowed simultaneously.
Within the applied coupling scheme we obtain 
$S^{\alpha\beta}(\mathbf{q},\omega_{n'n})+S^{\beta\alpha}(\mathbf{q},\omega_{n'n})=0$, 
$\forall \ n,n'$, $\alpha \ne\beta$ and $\sum_{\alpha} \Theta^{\alpha\alpha}=2$. 
The energy of the first experimental magnetic excitation is 
approximately $0.4$ meV. The corresponding peak is depicted on FIG. \ref{fig:NiExp}, 
see also~\cite{furrer_magnetic_PRB_2010}. 
This excitation is related with the transition between the ground state and the 
local singlet state $\lvert 0,1,1,\pm1 \rangle$. 
The associated scattering functions are
\begin{equation}\label{eq:ScatteringNikel12States1}
S^{\alpha \alpha}(\mathbf{q},\omega_{10}) = \tfrac{4}{9} 
[1-\cos(\mathbf{q}\cdot\mathbf{r}_{12})]p_0,
\ \ \
S^{zz}(\mathbf{q},\omega_{10}) = 0,
\end{equation}
where $p_n$ is the population factor and $\alpha = x,y$. 
The magnetic excitation at $0.6$ meV shown on FIG. \ref{fig:NiExp}
is associated with the eigenstate $\lvert 1,0,1,\pm1 \rangle$ 
and the scattering functions
\begin{equation}\label{eq:ScatteringNikel34States1}
S^{\alpha \alpha}(\mathbf{q},\omega_{30}) = 
\tfrac{4}{9} [1-\cos(\mathbf{q}\cdot\mathbf{r}_{34})]p_0,
\ \ \
S^{zz}(\mathbf{q},\omega_{30}) = 0,
\end{equation}
where $\alpha = x,y$. 
The functions
\eqref{eq:ScatteringNikel12States1} differ from
\eqref{eq:ScatteringNikel34States1} due to the spatial orientations
of the spin bonds with
$\mathbf{r}_{12}\cdot\mathbf{r}_{34} = 0$. 
For the same reason, we
deduce that the third cold peak at $1.7$ meV, see FIG. \ref{fig:NiExp}, 
is related with the transition between the ground state
and non magnetic triplet $\lvert 0,1,1,0 \rangle$.
For $\alpha = x,y$ the corresponding
scattering functions read
\begin{equation*}
S^{zz}(\mathbf{q},\omega_{50}) = 
\tfrac{4}{9} [1-\cos(\mathbf{q}\cdot\mathbf{r}_{12}) ]p_0,
\quad
S^{\alpha \alpha}(\mathbf{q},\omega_{50}) = 0.
\end{equation*}
The excited magnetic transition at around $1.2$ meV shown by green and red items 
on FIG. \ref{fig:NiExp} is nicely reproduced by the scattering functions
\begin{equation*}
S^{\alpha \alpha}(\mathbf{q},\omega_{64}) = 
\tfrac{2}{3} [1-\cos(\mathbf{q}\cdot\mathbf{r}_{12})] p_4,
\quad
S^{zz}(\mathbf{q},\omega_{64}) = 0,
\end{equation*}
where $\alpha = x,y$. The initial state is
given by the triplet state $\lvert 1,1,1,\pm1 \rangle$ with two triplet bonds
and the final one appears to be $\lvert 0,1,1,0 \rangle$.
Hence if the neutron scatters from the Ni3-Ni4 dimer, then we have
$\mathbf{q}\cdot\mathbf{r}_{12} = 0$ and
$\mathbf{q}\cdot\mathbf{r}_{34} > 0$. 
Nevertheless, with the coefficients $a^{s_{12},m_{12}}_{12}$ and 
$a^{s_{34},m_{34}}_{34}$ one can uniquely identify the 
two spin bonds and distinguish $I_{10}$ from $I_{30}$. 
Moreover, one can distinguish the eigenvalues
of tetramer Hamiltonian corresponding to $m=0$ and $m\ne0$, 
with $S^{zz}(\mathbf{q},\omega_{n'n}) = 0$ and 
$S^{xx}(\mathbf{q},\omega_{n'n})=0$, $S^{yy}(\mathbf{q},\omega_{n'n})=0$, 
respectively.
This affects directly the integrated intensities, such that choosing
$\mathbf{r}_{12} = (0,0,r^z)$ and $\mathbf{r}_{34} = (r^x,0,0)$ from
\eqref{eq:ScatteringIntensities} yields
\begin{equation*}\label{eq:NickelIntegIntensities}
\begin{array}{l}
\displaystyle
I_{10} \propto \gamma_{10}
\left[ 
1-\frac{\sin(\mathrm{q}\mathrm{r})}{\mathrm{q}\mathrm{r}}
\right]  F^2(\mathrm{q}),
\\ [0.5cm]
\displaystyle
I_{30} \propto \gamma_{30}
\left[ 
1 - 6\frac{\sin(\mathrm{q}\mathrm{r})}{5(\mathrm{q}\mathrm{r})^3} - 3\frac{\sin(\mathrm{q}\mathrm{r})}{5\mathrm{q}\mathrm{r}} + 6\frac{\cos(\mathrm{q}\mathrm{r})}{5(\mathrm{q}\mathrm{r})^2}
\right] F^2(\mathrm{q}),
\\[0.5cm]
\displaystyle
I_{50} \propto \gamma_{50} 
\left[ 
1-3\frac{\sin(\mathrm{q}\mathrm{r})}{(\mathrm{q}\mathrm{r})^3}+
3\frac{\cos(\mathrm{q}\mathrm{r})}{(\mathrm{q}\mathrm{r})^2}
\right] F^2(\mathrm{q}),
\\[0.5cm]
\displaystyle
I_{64} \propto \gamma_{64}
\left[ 
1-\frac{\sin(\mathrm{q}\mathrm{r})}{\mathrm{q}\mathrm{r}}
\right]  F^2(\mathrm{q}),
\end{array}
\end{equation*}
where 
$$
\gamma_{10} = \tfrac{8}{9}p_0, \ \gamma_{30} = \tfrac{20}{27}p_0, \ 
\gamma_{50} = \tfrac{8}{27}p_0, \ \gamma_{64} = \tfrac{4}{3}p_4,
$$
and $r=|\mathbf{r}_{12}|=|\mathbf{r}_{34}|$. 
The integrated intensities as a function of temperature
are shown on FIG. \ref{fig:NiIntensity}. According to 
Ref. \cite{furrer_magnetic_PRB_2010} the average distance between
Ni-Ni ions is $\mathrm{r}=6.68 \ \mathrm{\mathring{A}}$. 
The magnitude of the scattering vector is fixed at $\mathrm{q}=1 \
\mathrm{\mathring{A}}^{-1}$ and the calculated form factor
for Ni$^{2+}$ di-cations is 
$
F(\mathrm{q}) = 256(16+\mathrm{q}^2 \mathrm{r}^2_{\mathrm{o}})^{-2},
$
where $\mathrm{r}_{\mathrm{o}} = 0.529 \, \mathrm{\mathring{A}}$ is the Bohr radius.
The dependence of normalized intensities, $I_{n'n} \to
I_{n'n}/\gamma_{n'n}$, on the scattering vector is shown on FIG. \ref{fig:NiIntFFactor}.


\begin{table}[!ht]
	\caption{
		Calculated values of integrated intensities $I_{n'n}$ [arb. units] at
		temperatures 2.4, 9.3 and 23 K, shown on
		FIG. \ref{fig:NiIntensity} as blue squares, green circles and red triangles,
		respectively. 
	}
	\label{tab:NickelIntensThreeTemp}
	\begin{center}
		\begin{tabular}{lcccc}
			\hline\noalign{\smallskip}
			Transitions & I & II & III & IV  \\ 
			\noalign{\smallskip}\hline\noalign{\smallskip}
			Intensities & $I_{10}$ & $I_{30}$ & $I_{50}$ & $I_{64}$  \\
			\noalign{\smallskip}\hline\noalign{\smallskip}	
			2.4~[K] & 0.137(6) & 0.120(3) & 0.051(5) & 0.008(9)  \\
			9.3~[K] & 0.039(8) & 0.034(8) & 0.014(9) & 0.026(5)  \\
			23~~[K] & 0.019(5) & 0.017(1) & 0.007(3) & 0.021(1)  \\	
			\noalign{\smallskip}\hline
		\end{tabular}
	\end{center}
\end{table}	

\begin{table}[!h]
	\caption{
		Values of the coupling constants and the quantities 
		$c^1_{12}$, $c^2_{12}$ and $c_{34}$
		for all magnetic excitations with energies given in 
		eq. \eqref{eq:TransitionEnergies}. The results are obtained
		by taking into account the experimental data of 
		Ref. \cite{nehrkorn_inelastic_2010,furrer_magnetic_PRB_2010}.
	}
	\label{tab:Nickel}
	\begin{center}
		\begin{tabular}{lcccc}
			\hline\noalign{\smallskip}
			Transitions & I & II & III & IV  \\
			\noalign{\smallskip}\hline\noalign{\smallskip}
			$E_{n'n}$[meV] & $E_{10}$ & $E_{30}$ & $E_{50}$ & $E_{64}$ \\
			{} & 0.4 & 0.6 & 1.7 & 1.15  \\
			\noalign{\smallskip}\hline\noalign{\smallskip}
			$J_{\phantom{c^1_{12}}}$~[meV] & 0.325 & 0.325 & 0.325 & 0.325  \\
			$J_{c^1_{12}}$~[meV] & 0.372 & $-$ & 0.372 & $-$  \\
			$J_{c^2_{12}}$~[meV] & $-$ & $-$ & $-$ & 0.353  \\
			$J_{c_{34}}$~[meV] & $-$ & 0.334 & $-$ & $-$  \\
			$c^1_{12}$ & 1.1923 & $-$ & 1.1923 & $-$ \\
			$c^2_{12}$ & $-$ & $-$ & $-$ & 1.1153 \\
			$c_{34}$ & $-$ & 1.0384 & $-$ & $-$ \\ 
			\noalign{\smallskip}\hline
		\end{tabular}
	\end{center}
\end{table}	

\subsection{Energy of the magnetic transitions}
The energy transition $E_{ij}$ between $i$th and $j$th levels, 
corresponding to the calculated scattering intensities are
\begin{equation}\label{eq:TransitionEnergies}
\begin{array}{ll}
E_{10} = 6J-4Jc^1_{12},
&
E_{30} = 6J-4Jc_{34},
\\ [0.2cm]
E_{50} = 10J-4Jc^1_{12},
&
E_{64} = 8J-4Jc^2_{12}.
\end{array}
\end{equation}
From the last equations we can take advantage of one more constraint to
determine $J$, $E_{50}-E_{10} = 4J$. According to the
experimental data
\cite{nehrkorn_inelastic_2010,furrer_magnetic_PRB_2010} the ground
state magnetic excitations are grouped in two relatively broadened
peaks. The first peak is centered at about 0.5 meV and the second one at 1.7
meV. 
Furthermore, the first peak is composed of two subbands with energies
$E_{10}=0.4$ meV and $E_{30}=0.6$ meV. The width of the second peak can be
explained by the presence of an energy band, where the transition
energies are restricted in the region 1.6 meV to 1.8 meV. 
Therefore,
setting $E_{50} = 1.7$ meV we obtain $E_{50}-E_{10} = 1.3$ meV and $J
= 0.325$ meV. The computed energy
transitions are depicted on FIG. \ref{fig:NickelSpectra}. The
centers of both energy bands referring
to the value $c^{2}_{12} = 1.1153$ are shown by 
dashed lines. The energies of all transitions and
the corresponding parameters are given in table~\ref{tab:Nickel}.

\section{Conclusion}\label{sec:conclusion}
We propose a formalism for exploring the physical properties
of molecular magnets with non trivial bridging structure.
The underlying concept lies on the hypothesis
that due the cluster symmetry, as well as its shape,
size and the chemical structure that surrounds 
the magnetic ions, the spatial distribution of
valence electrons
is not unique leading to a variation of the relevant Coulomb, hopping 
and direct exchange energies.

Studying the INS spectra of the compound $\mathrm{Ni_4Mo_{12}}$ 
with the Hamiltonian \eqref{eq:AddHamiltonian} we were
able to derive a detailed picture for the neutron scattering 
intensities FIG. \ref{fig:NiIntensity} and FIG. \ref{fig:NiIntFFactor}.
Hamiltonian \eqref{eq:NickelHamilton} leads to
energy spectrum with two energy bands, 
shown in FIG. \ref{fig:NickelSpectra}. These bands are related to 
the fact that the tetramer cluster exhibits two distinguishable 
with respect to the coefficients $a^{s_{12},m_{12}}_{12}$ and 
$a^{s_{34},m_{34}}_{34}$ bonds. 
We ascribe this feature to the
difference in the chemical environment around Ni1-Ni2 and Ni3-Ni4 couples
that give rise to distinct spatial distributions of the valence electrons.
This allowed a unique identification of the magnetic 
excitations. Thereby, the obtained energy bands explain the width of second 
ground state peaks centered at 1.7 meV and the splitting of the first 
one centered at 0.5 meV.
In particular, for $s_{12} = 0$, $s_{34} = 0$ and $i=1,2$ 
we get $|c^i_{12}|>1$ and 
$|c_{34}|>1$, respectively. 
Besides, according to \eqref{eq:cij_a1} we have $J<J_{c^{i}_{12}}$ and $J<J_{c_{34}}$,
see table~\ref{tab:Nickel}.
These inequalities signals that the strength 
of the exchange is amplified.
Furthermore, the inequality $J_{c_{34}}<J_{c^{i}_{12}}$
indicates that most probably
the density of electrons along Ni3-Ni4 bond is lower than that along the Ni1-Ni2.

To conclude we would like to anticipate that the results for the magnetization and
low-field magnetic field susceptibility of the tetramer $\mathrm{Ni_4Mo_{12}}$ are in
concert with the experimental measurements. These results will be the
subject of a separate paper.

\section*{Acknowledgment}
The authors are indebted to Prof. N.S. Tonchev and Prof. N. Ivanov
for the very helpful discussions and to Prof. J. Schnack for providing
us the experimental results on FIG. \ref{fig:NiExp}. 
This work was supported by the Bulgarian National Science Fund under
contract DN/08/18.


%

\end{document}